\documentclass[prd,showpacs,twocolumn,superscriptaddress,floatfix,nofootinbib,10pt]{revtex4-2}
\usepackage{setspace}
\usepackage[utf8]{inputenc}
\usepackage{float}
\usepackage{amsmath,amssymb,amsfonts,bm}
\usepackage{graphicx}
\usepackage{multirow}
\usepackage[usenames,dvipsnames]{color}
\allowdisplaybreaks
\usepackage[
colorlinks=true,
linkcolor=blue,
breaklinks=true,
urlcolor=blue,
citecolor=blue]{hyperref}

\usepackage{orcidlink}
\usepackage{subfigure}
\usepackage{soul}
\usepackage{xcolor}
\usepackage{physics}
\usepackage{booktabs}
\usepackage{ulem}
\usepackage{tikz-feynman}

\definecolor{green}{rgb}{0,0.6,0}
\newcommand{\mev}{\textrm{ MeV}}

\newcommand{\GXNU}{\affiliation{Department of Physics, Guangxi Normal University, Guilin 541004, China}}

\newcommand{\GXZD}{\affiliation{Guangxi Key Laboratory of Nuclear Physics and Technology, Guangxi Normal University, Guilin 541004, China}}

\newcommand{\SEU}{\affiliation{School of Physics, Southeast University, Nanjing 210094, China}}

\newcommand{\IFIC}{\affiliation{Departamento de F\'{\i}sica Te\'orica and IFIC, Centro Mixto Universidad de
Valencia-CSIC Institutos de Investigaci\'on de Paterna, Apartado 22085,
46071 Valencia, Spain}}

\begin{document}
\title{Determination of the $K^{+}\bar{K}^{0}$ scattering length and effective range and its relation to the $a_{0}^{+}(980)$ from the $\chi_{c1}\to\pi^{+}\pi^{-}\eta$ reaction}

\begin{abstract}
	We analyze the clean cusp, seen in the $\eta \pi$ mass distribution with high precision of the $\chi_{c1} \to \eta \pi^+ \pi^-$ reaction in the BESIII experiment, with the aim of making a precise determination of the scattering length $a$ and effective range $r_0$ of $K^+ \bar{K}^{0}$. For that, we follow a previous theoretical work that gave a good reproduction of these data using the chiral unitary approach for the meson-meson interaction, and allow some flexibility in the input to carry a better fit to the data. The important task of determining the uncertainties in the scattering parameters is done using the resampling method and an accuracy in $a$ and $r_0$ is obtained better than $20\%$. The effective range is determined for the first time with this analysis.
\end{abstract}

\author{Hai-Peng Li}%
\GXNU%
\author{Jia-Xin Lin}%
\SEU%
\author{Wei-Hong Liang\orcidlink{0000-0001-5847-2498}}%
\email{liangwh@gxnu.edu.cn}
\GXNU%
\GXZD%
%
\author{Eulogio Oset}%
\email{Oset@ific.uv.es}
\GXNU%
\IFIC%

\maketitle

\section{Introduction}\label{sec:Intr}
The determination of scattering lengths and effective ranges from cusps seen in some reaction has proved effective, as shown for instance in the determination of the $\pi\pi$ scattering length from the $K$ decay into pions \cite{Colangelo2006,Batley2009}.
More recently, it was found that the high precision Belle data on the $\Lambda_{c}^{+}\to pK^{-}\pi^{+}$ reaction \cite{Yang2023}, which shows a cusp at the $\eta\Lambda$ threshold in the $K^{-}p$ invariant mass distribution, served to determine the $\eta\Lambda$ scattering length and effective range, as well as the position of the $\Lambda(1670)$ resonance, with an unprecedented precision \cite{Duan2024}.
Motivated by this success, we wish to exploit the idea to obtain the $K^{+}\bar{K}^{0}$ scattering length and effective range, as well as information on the $a_{0}(980)$ resonance, for the high resolution BESIII experiment on the $\chi_{c1}\to\eta\pi^{+}\pi^{-}$ reaction~\cite{Ablikim2017}, which shows a clear cusp for the $\eta\pi^{+}$, $\eta\pi^{-}$ invariant mass distribution at the $K^{+}\bar{K}^{0}$ threshold.
Similarly to the case of Ref.~\cite{Duan2024}, where the $\eta\Lambda$ scattering parameters are determined from a peak at the $\eta\Lambda $ threshold of the $K^{-}p$ invariant mass distribution, in the precent case we shall ditermine the scatteringg parameters of the $K^{+}\bar{K}^{0}$ system from a peak of the $\eta\pi^{+}$, $\eta\pi^{-}$ invariant mass distribution at the $K^{+}\bar{K}^{0}$ threshold.
One should stress that in both cases we determine the scattering parameters of one channel from one reaction in which this channel is not measured.
The particles observed are different, but correspond to pairs with the same quantum numbers, which necessarily couple to the channel investigated in any unitary approach.
This feature is what makes it possible to determine the scattering parameters of one channel from the observation of a cusp of a second channel at the threshold of the first one.
It looks at a first sight that such a task would have large uncertainties.
However, as demonstrated in Ref.~\cite{Duan2024}, the constraints of unitarity in coupled channels are so strong, that not only can one obtain information on the scattering parameters, but can determine them with high precision.

The determination of the $K^{+}\bar{K}^{0}$ scattering length has attracted much attention.
In Table~\ref{tab:tab1} we show the results obtained from the analysis of different experiments done in Ref.~\cite{Buescher2006}.
\begin{table*}[htbp]
\caption{Result for the $K^{+}\bar{K}^{0}$ scattering length from different works.}
\label{tab:tab1}
\begin{ruledtabular}
\begin{tabular}{cccccc}
 References & Ref.~\cite{Aloisio2002}&Ref.~\cite{Achasov2003} &Ref.~\cite{Achasov2000}  & Ref.~\cite{Bugg1994}  & Ref.~\cite{Buescher2006} \\
\colrule
$a_1$ (fm) & $+0.071-0.66i$ & $-0.006-0.76i$ & $-0.13-2.23i$ & $-0.075-0.70i$ & $(-0.02\pm 0.02)-(0.63\pm 0.24)i$\\
\colrule
\colrule
References& Ref.~\cite{Achasov2003}   & Ref.~\cite{Abele1998}  & Ref.~\cite{Teige1999}  & Ref.~\cite{Bargiotti2003} \\
\colrule
$a_1$ (fm) & $-0.16-1.05i$ & $-0.13-0.61i$ & $-0.16-0.59i$ & $-0.54-1.89i$\\
\end{tabular}
\end{ruledtabular}
\end{table*}
In Refs.~\cite{Aloisio2002,Achasov2003,Achasov2000,Bugg1994} the information is obtained from the $\phi\to\eta\pi^{0}\gamma$ reaction and the extracted value of the coupling of the $a_{0}(980)$ resonance to $K\bar{K}$.
The same is done from Ref.~\cite{Teige1999} in the study of the $\pi^{-}p\to\eta\pi^{+}\pi^{-}n$ reaction.
In Refs.~\cite{Bargiotti2003,Abele1998} the reactions $\bar{p}p\to K^{+}K^{0}_{S}\pi^{-}, K^{+}K^{0}_{L}\pi^{+}$ are measured, and, once again, the scattering lenght is obtained in Ref.~\cite{Buescher2006} from the values of the coupling of the $a_{0}(980)$ to $K^{+}\bar{K}^{0}$.

It should be stated from the beginning that the concept of the coupling of $a_{0}(980)$ to $K\bar{K}$ is problematic if the $a_{0}(980)$ does not correspond to a bound state of $K\bar{K}$, or in other words, if there is not a pole for this state in the ordinary second Riemann sheet (corresponding to a virtual state in a different notation).
A parameterization of the $K^{+}\bar{K}^{0}$ amplitude is then also problematic experimentally when one has a cusp, since amplitudes close to threshold have a very peculiar behavior \cite{Matuschek2021,Zhou2015,Dong2021,Zhang2024}.

In Ref.~\cite{Buescher2006}, the reaction $pp\to dK^{+}\bar{K}^{0}$ near threshold is used to determine the $K^{+}\bar{K}^{0}$ scattering length, but, as shown in Ref.~\cite{Oset2001}, this fusion reaction is complicated, leading to uncertainties in the determination of that scattering length.

As we can see in Table~\ref{tab:tab1}, there is a large dispersion of the results obtained from the different analyses.
A more precise determination, with its uncertainty, should be more welcome, and this is the purpose of the present work.
In addition, we also determine the effective range for the first time.
We, thus, study the BESIII $\chi_{c1}\to \eta\pi^{+}\pi^{-}$ reaction \cite{Ablikim2017} and use the approach of Ref.~\cite{Liang2016}, which was shown to give a good reproduction of the experimental results using input of the chiral unitary approach for the meson-meson interactions \cite{Oller2000}.
Yet, what we do is to leave freedom to the input parameters and carry a fit to the data to determine them, then the framework provides the $K^{+}\bar{K}^{0}$ scattering length and effective range.
In other words, what we would be doing is a fit to the data using basically a model independent (free meson-meson transition potentials) procedure, with the only constraints of unitarity in coupled channels.
The determination of the errors of $a$ and $r_{0}$ is then accomplished by the method of resampling \cite{Press1992,Efron1986,Albaladejo2016} generating random Gaussian weighed centroids for all the data and carrying a fit in each case.
With about $50$ different fits, in which the parameters of the theory are determined, the value of $a$ and $r_{0}$ are calculated and the average and dispersion of these values are then evaluated.
The method proves efficient when there are correlations between the different parameters.
Hence, the values of the parameters in each fit can change, but the results obtained for $a$ and $r_{0}$ are stable \cite{Duan2024,Molina2024,Feijoo2024,Albaladejo2023,Ikeno2023}.

\section{formalism}
In the $\chi_{c1}\to\eta\pi^{+}\pi^{-}$ reaction, the $\chi_{c1}$ state has $I^{G}(J^{PC})=0^{+}(1^{++})$.
The $\eta,\pi^{+},\pi^{-}$ have $J^{P}=0^{-}$.
Since the decay proceeds via strong interaction, conservation of spin parity implies that the reaction proceeds in $P$-wave.
The amplitude must have the form of $\vec{\epsilon}_{\chi_{c1}}\cdot \vec{p}_{i}$ and symmetry over the three mesons implies the struction of the decay amplitude as
\begin{equation}
	\label{eq:t}
	t=A\left(\vec{\epsilon}_{\chi_{c1}}\cdot \vec{p}_{\eta}+\vec{\epsilon}_{\chi_{c1}}\cdot \vec{p}_{\pi^{+}}+\vec{\epsilon}_{\chi_{c1}}\cdot \vec{p}_{\pi^{-}}\right).
\end{equation}

The next step is to consider that $\chi_{c1}$ is a singlet of SU(3) and assume that we have SU(3) symmetry in the $\chi_{c1}\to P_{1}P_{2}P_{3}$ transition at the tree level, where $P_{1},P_{2}$ and $P_{3}$ are three pseudoscalar mesons.
One might argue that there can be some SU(3) breaking, but at this point it is interesting to mention that in the $P_{1}P_{2}$ meson interaction, the chiral lagrangians are SU(3) symmetric \cite{Gasser1984,Scherer2003}, and SU(3) symmetry is broken due to loops in the unitary treatment of the interaction, where the different masses of the particles in the same SU(3) multiplet have important effects \cite{Jido2003}.
Hence, the assumption of SU(3) symmetry at the tree level is a good starting point, and the success of Ref.~\cite{Liang2016} to reproduce the spectra in the $\chi_{c1}\to \eta\pi^{+}\pi^{-}$ reaction proved it.

This means that in the structure of $t$ in Eq. \eqref{eq:t}, one must consider all terms of $\eta P_{1}P_{2}$ structure for the first term and then allow $P_{1}P_{2}$ to interact to give $\pi^{+}\pi^{-}$, or $\pi^{+} P_{3}P_{4}$ in the second term and allow $P_{3}P_{4}$ interact to give $\eta\pi^{-}$, or $\pi^{-}P_{5}P_{6}$ in the third term and allow $P_{5}P_{6}$ interact to give $\eta\pi^{+}$ in the final states.
Even then, there are three possible SU(3) singlet structures made from the $\mathcal{P}\equiv q_{i}\bar{q}_{j}$ matrix ($q_{i}=u,d,s$ quarks), where $\mathcal{P}$ in terms of pseudoscalars is given by
\begin{align}
	\nonumber
	&\mathcal{P}\equiv\\
	&\resizebox{\columnwidth}{!}{$
	\begin{pmatrix}
	\frac{1}{\sqrt{2}}\pi^{0}+\frac{1}{\sqrt{3}}\eta+\frac{1}{\sqrt{6}}\eta'&\pi^{+}&K^{+}\\
	\pi^{-}&-\frac{1}{\sqrt{2}}\pi^{0}+\frac{1}{\sqrt{3}}\eta+\frac{1}{\sqrt{6}}\eta'&K^{0}\\
	K^{-}&\bar{K}^{0}&-\frac{1}{\sqrt{3}}\eta+\sqrt{\frac{2}{3}}\eta'\\
	\end{pmatrix}
   $},
\end{align}
which implies the standard $\eta-\eta'$ mixing of Ref.~\cite{Bramon1992}.
The SU(3) singlet structures are $\langle \mathcal{PPP}\rangle$, $\langle \mathcal{P}\rangle\langle \mathcal{PP}\rangle$ and $\langle \mathcal{P}\rangle\langle \mathcal{P}\rangle\langle \mathcal{P}\rangle$, where $\langle \cdots\rangle$ indicates the trace in SU(3) of these matrices.
Yet, the structure with less traces, $\langle \mathcal{PPP}\rangle$, is dominant according to Ref.~\cite{Manohar1998}, and in Ref.~\cite{Debastiani2017} it was shown that the $\langle \mathcal{P}\rangle\langle \mathcal{PP}\rangle$ structure led to disastious results in the mass distributions.

\begin{figure}[htbp]
	\begin{center}
	\includegraphics[scale=0.7]{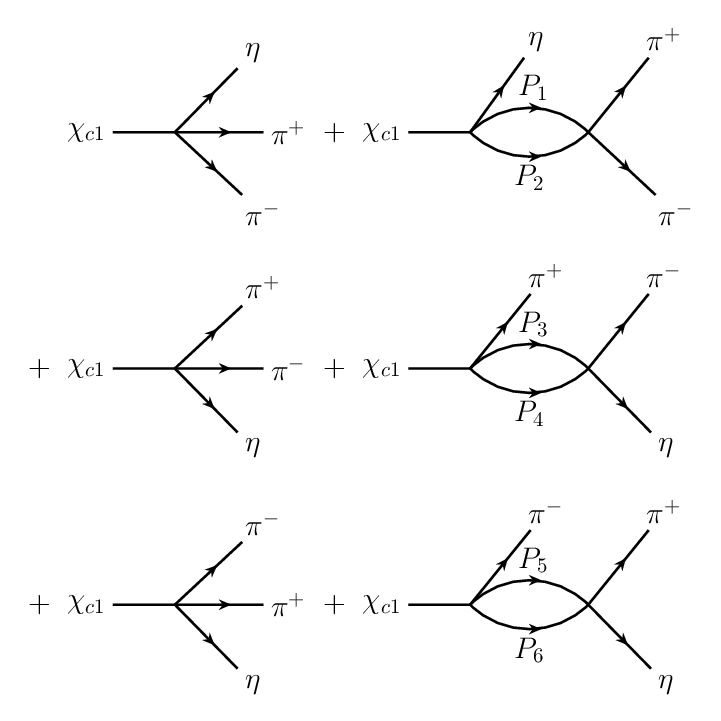}
	\end{center}
	\vspace{-0.7cm}
	\caption{Diagrams entering the transition $\chi_{c1}\to\eta\pi^{+}\pi^{-}$.}
	\label{fig:feyn}
\end{figure}

With all these supporting grounds, it was found in Ref.~\cite{Liang2016} that the terms of type $\eta P_{1}P_{2}$ were given by
\begin{equation}
	C_{1}:\quad \eta \left( \frac{6}{\sqrt{3}}\pi^{+}\pi^{-}+\frac{3}{\sqrt{3}}\pi^{0}\pi^{0}+\frac{1}{3\sqrt{3}}\eta\eta \right),
\end{equation}
those of $\pi^{+}P_{3}P_{4}$ by
\begin{equation}
	C_{2}:\quad \pi^{+}\left( \frac{6}{\sqrt{3}}\pi^{-}\eta+3K^{0}K^{-}\right),
\end{equation}
and those of $\pi^{-}P_{5}P_{6}$ by
\begin{equation}
	C_{3}:\quad \pi^{-}\left( \frac{6}{\sqrt{3}}\pi^{+}\eta+3K^{+}\bar{K}^{0}\right),
\end{equation}
and then, diagrammatically, the transition is produced by the terms of Fig.~\ref{fig:feyn}, where the upper line in each diagram is the one carrying the momentum $\vec{p}_{i}$ in Eq.~\eqref{eq:t}.

Analytically, the diagrams in Fig.~\ref{fig:feyn} transform in the amplitude
\begin{equation} \label{eq:amplitude}
	t=t_{\eta}+t_{\pi^{+}}+t_{\pi^{-}},
\end{equation}
where
\begin{equation}
t_{\eta}=\left(\vec{\epsilon}_{\chi_{c1}}\cdot \vec{p}_{\eta}\right)\tilde{t}_{\eta},
\end{equation}
with
\begin{equation}
	\label{eq:teta}
	\tilde{t}_{\eta}=V_{P}\left(h_{\pi^{+}\pi^{-}}+\sum_{i}h_{i}\,S_{i}\,G_{i}(M_{\text{inv}})\;t_{i,\pi^{+}\pi^{-}}\right),
\end{equation}
where
\begin{equation}
	h_{\pi^{+}\pi^{-}}=\frac{6}{\sqrt{3}},\quad h_{\pi^{0}\pi^{0}}=\frac{3}{\sqrt{3}},\quad h_{\eta\eta}=\frac{1}{3\sqrt{3}},
\end{equation}
and $S_{i}$ are symmetry factors for the identical particles, given by \cite{Liang2016}
\begin{equation}
	S_{\pi^{0}\pi^{0}}=1;\quad S_{\eta\eta}=3.
\end{equation}
The second term $t_{\pi^{+}}$ of Eq.~\eqref{eq:amplitude} is given by
\begin{equation}
	t_{\pi^{+}}=\left(\vec{\epsilon}_{\chi_{c1}}\cdot \vec{p}_{\pi^{+}}\right)\tilde{t}_{\pi^{+}},
\end{equation}
with
\begin{equation}
	\label{eq:tpip}
	\tilde{t}_{\pi^{+}}=V_{P}\left(h_{\pi^{-}\eta}+\sum_{i}h_{i}\, G_{i}(M_{\text{inv}})\;t_{i,\pi^{-}\eta}\right),
\end{equation}
and
\begin{equation}
	h_{\pi^{-}\eta}=\frac{6}{\sqrt{3}},\quad h_{K^{0}K^{-}}=3.
\end{equation}
The third term $t_{\pi^{-}}$ of Eq.~\eqref{eq:amplitude} is given by
\begin{equation}
	t_{\pi^{-}}=\left(\vec{\epsilon}_{\chi_{c1}}\cdot \vec{p}_{\pi^{-}}\right)\tilde{t}_{\pi^{-}},
\end{equation}
with
\begin{equation}
	\label{eq:tpim}
	\tilde{t}_{\pi^{-}}=V_{P}\left(h_{\pi^{+}\eta}+\sum_{i}h_{i}\, G_{i}(M_{\text{inv}}) \; t_{i,\pi^{+}\eta}\right),
\end{equation}
and
\begin{equation}
	h_{\pi^{+}\eta}=\frac{6}{\sqrt{3}},\quad h_{\bar{K}^{0}K^{+}}=3.
\end{equation}
The factor $V_{P}$ is an unknown normalization constant, which is linked to the unnormalized mass distribution of Ref.~\cite{Ablikim2017}.

The mass distributions are obtained by the master formula of the PDG \cite{Navas2024}
\begin{equation}
	\label{eq:gamma}
	\frac{\dd^2\Gamma}{\dd M_{12} \, \dd M_{23}}=
	\frac{1}{(2\pi)^3}\,\frac{1}{8M^3_{\chi_{c1}}}M_{12}\, M_{23}\overline\sum\sum\abs{t}^2,
\end{equation}
and the single mass distribution are obtained by
\begin{equation}
	\label{eq:gamma2}
	\frac{\dd \Gamma}{\dd M_{12}}=\int\frac{\dd^2\Gamma}{\dd M_{12}\,\dd M_{23}}\; \dd M_{23},
\end{equation}
with the limits of Ref.~\cite{Navas2024} and appropriate permutations of the particle indices.
Note also that, as shown in Ref.~\cite{Liang2016}, there is no interference between the terms of Eq.~\eqref{eq:t} in $\abs{t}^2$, which also allows to directly calculate ${\dd\Gamma}/{\dd M_{ij}}$ without the integral of Eq.~\eqref{eq:gamma2} as shown in Ref.~\cite{Liang2016}.

In the $\pi^+\pi^-$ invariant mass distribution from BESIII data \cite{Ablikim2017}, the $f_2(1270)$ state shows up at the region of $[1, 1.5] \, \rm GeV$. 
We take a Breit-Wigner shape of $\dfrac{A}{M^2_{\text{inv}}(\pi\pi)-m_{f_{2}}^2+i m_{f_{2}}\Gamma_{f_{2}}}$ to account for the tail of the $f_2(1270)$, with $m_{f_{2}}$ and $\Gamma_{f_{2}}$ the physical mass and width of the $f_2(1270)$.
In order to compare with the $\pi^+\pi^-$ data, we follow Ref.~\cite{Liang2016} and add a background coming from the $a_0(980)$ peak, which is taken linear in the $\pi\pi$ invariant mass, $B\,[M_{\text{inv}}(\pi\pi)-2m_{\pi}]$.

\section{determination of $a$ and $r_{0}$ for $K^{+}\bar{K}^{0}$}

In order to evaluate the $t$ matrices entering Eqs.~\eqref{eq:teta}, \eqref{eq:tpip} and \eqref{eq:tpim}, we use the formulas of $V_{ij}$ from Refs.~\cite{Liang2014,Xie2015}, which are calculated  for pairs of neutral charge and use
\begin{align}\label{eq:t_structure}
    V_{\pi^+ \eta, \,\pi^+ \eta} &= V_{\pi^0 \eta, \,\pi^0 \eta}, \nonumber \\
	V_{K^0 K^-, \,\pi^- \eta} &= \sqrt{2} \,V_{K^+ K^-, \,\pi^0 \eta},  \\
	V_{K^+ \bar K^0, \,K^+ \bar K^0} &= \frac{1}{2}\left(V_{K^0 \bar K^0, \,K^0 \bar K^0} + V_{K^+K^-, \,K^+ K^-} \right. \nonumber\\
	& \quad ~~~\left.-2\, V_{K^+K^-, \,K^0 \bar K^0}\right). \nonumber
\end{align}
Then, the $T$ matrices are obtained from the Bethe-Salpeter equation in coupled channels
\begin{equation}
	T=[1-VG]^{-1}V,
\end{equation}
where $G$ is the loop function of pairs of pseudoscalars, which in the cut off regularization is given by
\begin{align}
	\nonumber
	G_{i}(s)=\int_{\abs{\vec{q\,}}<q_{\text{max}}}&\frac{\dd^3q}{(2\pi)^3}\;\frac{\omega_{1}(q)+\omega_{2}(q)}{2\,\omega_{1}(q)\;\omega_{2}(q)}\\
	&\times\frac{1}{s-[\omega_{1}(q)+\omega_{2}(q)]^2+i\epsilon},
\end{align}
with the subindices $1, 2$ refer the two mesons in channel $i$, and $\omega_{l}=\sqrt{q^2+m_{l}^2}$.
A value of $q_{\text{max}}=600-630 \mev$ was used in Ref.~\cite{Liang2016}, with small difference between the two options.
In order to avoid using the $t$ matrices of the chiral unitary approach beyond the region of validity, we cut them adiabatically, as done in Ref.~\cite{Debastiani2017} by using
\begin{equation}
	G\,t(M_{\text{inv}})=G\,t(M_{\text{cut}})\;\text{e}^{-\alpha(M_{\text{inv}}-M_{\text{cut}})}, ~~~ \text{for} ~ M_{\text{inv}}>M_{\text{cut}},
\end{equation}
with $M_{\text{cut}}=1100 \mev$, $\alpha=0.0054 \mev^{-1}$, and it was found in Ref.~\cite{Debastiani2017} that the changes in the mass dirtributions induced by moderate changes of $\alpha$ and $M_{\text{cut}}$, were minor.

The scattering length $a$ and effective range $r_{0}$
are given via Ref.~\cite{Ikeno2023} as
\begin{align}
	\nonumber
	-\frac{1}{a}&=-8\pi\,\sqrt{s}\;T^{-1}\Big|_{s=s_{\rm th}},\\
	r_{0}&=\frac{\sqrt{s}}{\mu}\; \frac{\partial}{\partial s} \;2\left(-8\pi \,\sqrt{s} \;T^{-1}+ik\right)\Big|_{s=s_{\rm th}},
\end{align}
with
\begin{equation}
	k=\frac{\lambda^{1/2}(s,m_{1}^2,m_{2}^2)}{2\sqrt{s}},
\end{equation}
where $s_{\rm th}$ is the squared of the energy of the system at threshold and $\mu$ is the reduced mass of $m_{1}$ and $m_{2}$.
\section{results}
The first result that we want to show is the value of $a$ and $r_{0}$, with the chiral unitary input, using the potential of Refs.~\cite{Liang2014,Xie2015} with $f=93$ MeV, and different values of $q_{\text{max}}$.
This is shown in Table~\ref{tab:1}.
\begin{table}[!b]
\centering
\caption{The observables of the chiral unitary approach with $f=93$ MeV.}
\label{tab:1}
\setlength{\tabcolsep}{10pt}
\begin{tabular}{ccc}
\toprule
 $q_{\text{max}}$(MeV)&$a_{K^+ \bar{K}^0}$(fm) & $r_{0,K^+ \bar{K}^0}$(fm) \\
\midrule
	600&$-0.191-i0.674$ & $-0.917-i0.237$ \\
	\hline
	630&$-0.166-i0.727$ & $-0.897-i0.198$ \\
	\hline
	740& $0.004-i0.892$ & $-0.855-i0.105$\\
\bottomrule
\end{tabular}
\end{table}
We can see that the results for $\Re [a_{K^{+}\bar{K}^{0}}]$ change much by changing $q_{\text{max}}$.
However for $q_{\text{max}}=600-630 \mev$, which are our favorite choices (see Ref.~\cite{Debastiani2017}),
$\Re [a_{K^{+}\bar{K}^{0}}]$ are stable between $-0.19$ and $-0.17$ fm.
The values of $\Im [a_{K^{+}\bar{K}^{0}}]$ are more stable, and in the range of $q_{\text{max}}\in[600,630] \mev$ they change between $-0.73$ and $-0.67$ fm.
The values of $r_{0}$ are also rather stable ranging in $\Re [r_{0}]\in[-0.90,-0.92]$ fm and $\Im [r_{0}] \in[-0.20,-0.24]$ fm in the same range of $q_{\text{max}}$.

The other point we want to stress is that we do not find a pole for the $a_{0}(980)$ in the range of $q_{\text{max}}$ of Table \ref{tab:1} with the usual prescription for the second Riemann sheet of taking
\begin{equation}
	G_i^{II}(s)=G_i(s)+\frac{i}{4\pi\sqrt{s}}q_{\text{on}};\quad \text{for}\ \Re \sqrt{s}>\sqrt{s}_{\rm th},
\end{equation}
with $q_{\text{on}}=\lambda^{1/2}(s,m_{1}^2,m_{2}^2)/2\sqrt{s}$, $\Im(q_{\text{on}})>0$, and $\sqrt{s}_{\rm th}$ the threshold mass of the corresponding channel.
Thus, the $a_{0}(980)$ would qualify as a cusp of the $K^{+}\bar{K}^{0}$ threshold, or in other words, a barely failed bound state, or virtual state.
We would have to go to values of $q_{\text{max}}>1000 \mev$ to get a pole, but this would greatly distort the agreement of the theory of Ref.~\cite{Liang2016} with the experiment of Ref.~\cite{Ablikim2017}.

\subsection{Resampling to get $a$, $r_{0}$ with errors}
Next we resort to the resampling method to determine $a$, $r_{0}$ with their respective uncertainties.
For this we do the following.
All terms of $V_{ij}$ in Refs.~\cite{Liang2014,Xie2015} are proportional to $\frac{1}{f^2}$, where $f$ is the pion decay constant, taken in Ref.~\cite{Liang2016} as $f=93 \mev$.
Here, in order to have freedom in the potential and carry a fair model independent analysis, we change
\begin{align}
	\nonumber
	f\to& f_{\pi\eta},& \text{in the $\pi\eta$ channel};\\
	\nonumber
	f\to& f_{\pi\pi},& \text{in the $\pi\pi$ channel};\\
	\nonumber
	f\to& f_{K\bar{K}},& \text{in the $K\bar{K}$ channel};\\
	\nonumber
	f\to& f_{\eta\eta},& \text{in the $\eta\eta$ channel};
\end{align}
with $f_{i}\in[40,180] \mev$ and $f_{K^+K^-}=f_{K^0 \bar K^0}$.
The last equation is taken to preserve isospin invariance in the potential.
\footnote{The $\pi\pi$ and $\eta\eta$ channels are not strictly necessary here if we are concerned just around the $a_{0}(980)$ peak with $I=1$.
However, since in Refs.~\cite{Liang2014,Xie2015} all channels are considered simultaneously, we consider them here too, and they are needed to get the $\pi^{+}\pi^{-}$ distribution at the same time.}

We are also flexible with the $q_{\text{max}}$ parameter, which we take in the range $q_{\text{max}}\in[400,1500] \mev$.
In Table~\ref{tab:2} we give the results for the $f_{i}$ parameters and $a_{K^{+}\bar{K}^{0}}$, $r_{0,K^{+}\bar{K}^{0}}$ and $q_{\text{max}}$.
\begin{table*}[htbp]
\centering
\caption{The results of fitting the resampled data.}
\label{tab:2}
\setlength{\tabcolsep}{22pt}
\begin{tabular}{ccc}
\toprule
$f_{\pi\pi}$ & $f_{KK}$ & $f_{\eta\eta}$ \\
\midrule
$100.301\pm1.521$ & $101.510\pm 4.130$ & $83.511\pm13.761$ \\
\bottomrule
 $f_{\pi\eta}$& $q_{\text{max}}$&\\
\midrule
$117.836\pm7.770$&$809.327\pm22.217$ \\
\bottomrule
\end{tabular}\\[0.3cm]
\setlength{\tabcolsep}{10pt}
\begin{tabular}{cc}
\toprule
   $a_{K^+ \bar{K}^0}$(fm) & $r_{0,K^+ \bar{K}^0}$(fm) \\
\midrule
  $-(0.371\pm 0.045)-i(0.549\pm0.023)$ & $-(0.982\pm 0.107)-i(0.265\pm0.035)$ \\
\bottomrule
\end{tabular}
\end{table*}
We should stress that, since there are correlations in the parameters, particularly with $q_{\text{max}}$, we should not pay excessive attention to the values of the parameters, but just to $a$ and $r_{0}$.
We observe that $\Re a\approx -0.41$ fm with an uncertainty of about $18\%$, and $\Im a\approx -0.52$ fm with an uncertainty of about $5\%$.
This reflects the results obtained in Table~\ref{tab:1} where the dispersion among the values of $\Re a$ is large, while the dispersion in $\Im a$ is smaller, indicating that $\Im a$ can be obtained with more precision than $\Re a$.
Concerning the effective range, $r_{0}$, we obtain $\Re r_{0}\approx-0.85$ fm with a precision of $18\%$, and $\Im r_{0}\approx-0.32$ fm with a precision of about $11\%$.
We should note that this is the first time that values for $r_{0}$ are provided, to the best of our knowledge.
The value obtained for $a$ with its uncertainty provide a reliable estimate of this magnitude, which is to be appreciated in view of the large dispersion of results in Table~\ref{tab:tab1}.

\begin{figure}[bp]
	\begin{center}
	\includegraphics[scale=0.53]{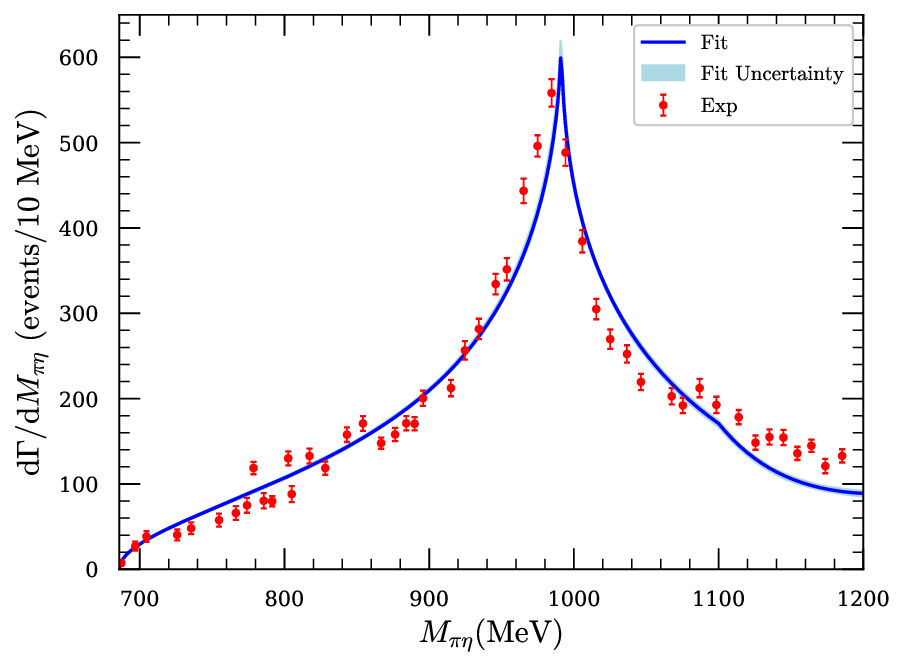}
	\end{center}
	\vspace{-0.7cm}
	\caption{(Color online) Results of the fits to $\pi\eta$ invariant mass distribution for the $\chi_{c1}\to\eta\pi^{+}\pi^{-}$ decay.}
	\label{fig:pe}
\end{figure}

\begin{figure}[tbp]
	\begin{center}
	\includegraphics[scale=0.53]{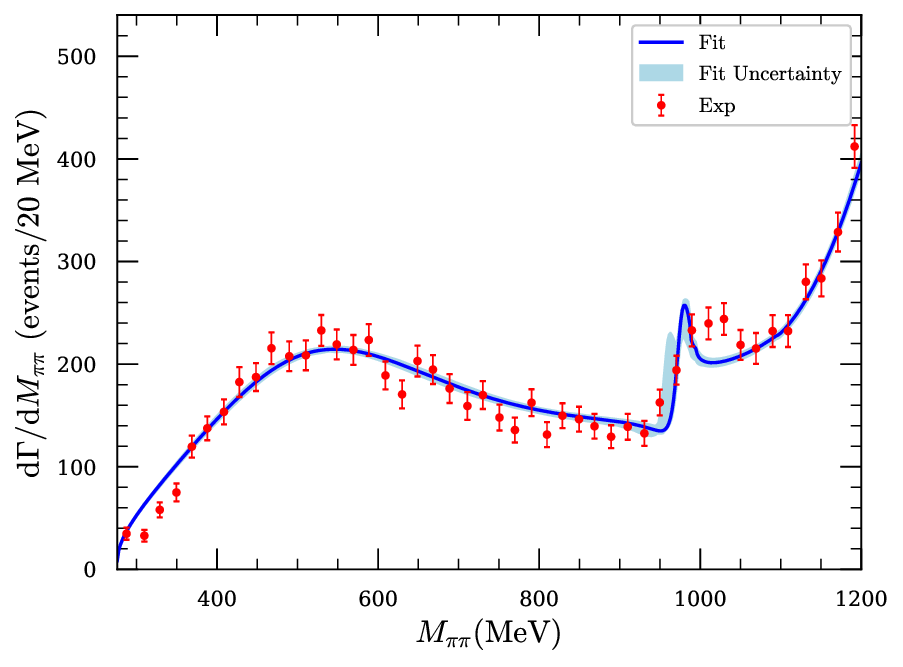}
	\end{center}
	\vspace{-0.7cm}
	\caption{(Color online) Results of the fits to $\pi\pi$ invariant mass distribution for the $\chi_{c1}\to\eta\pi^{+}\pi^{-}$ decay.}
	\label{fig:pp}
\end{figure}

We performed a simultaneous fit to both the $\pi\eta$ and $\pi\pi$ invariant mass distributions.
In Figs.~\ref{fig:pe} and \ref{fig:pp}, we show the results that we obtain for the $\eta\pi$ mass distribution and the $\pi\pi$ mass distribution respectively, compared with the data of Ref.~\cite{Ablikim2017}.
As we can see, the agreement with data is very good, and improves over the results obtained in Ref.~\cite{Liang2016} (see Fig.~6 of that reference), where one had no freedom in the parameters of the potential of the chiral unitary approach.

\begin{figure}[tp]
	\begin{center}
	\includegraphics[scale=0.51]{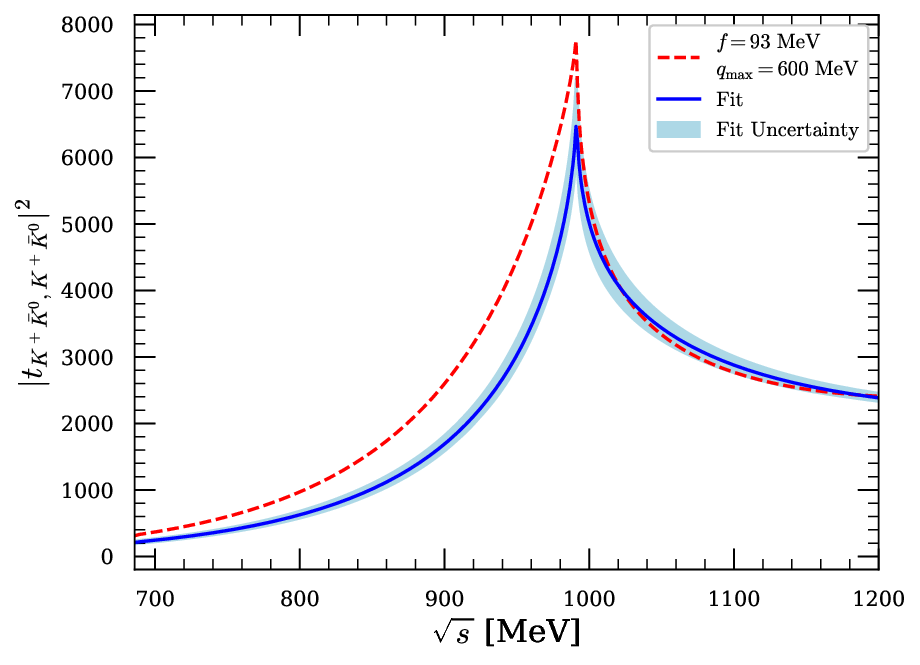}
	\end{center}
	\vspace{-0.7cm}
	\caption{(Color online) Absolute squared values of $t$ in $K^{+}\bar{K}^{0}\to K^{+}\bar{K}^{0}$ as a function of the $K^{+}\bar{K}^{0}$ invariant mass ($\sqrt{s}$). The deshed red line corresponds to the results of the chiral unitary approach with $q_{\text{max}}=600 \mev$. The blue solid line and the band correspond to the present fit with its uncertainty.}
	\label{fig:t}
\end{figure}
Next, in Fig.~\ref{fig:t} we would like to show the result for $\abs{t_{K^{+}\bar{K}^{0},\,K^{+}\bar{K}^{0}}}^2$ as a function of the $K^{+}\bar{K}^{0}$ invariant mass, with the dispersion given by the resampling procedure.
As is common, the band includes the $68\%$ of events closer to the average value.
As we can see, there are some small differences between the standard chiral unitary approach result and the new fit to the data with its uncertainty shown by the blue band.

\section{Conclusions}
We have studied the $\chi_{c1}\to\eta\pi^{+}\pi^{-}$ reaction, where a clean cusp with high resolution around the $K^{+}\bar{K}^{0}$ threshold is seen in the $\eta\pi$ mass distribution, with the aim of determining the $K^{+}\bar{K}^{0}$ scattering length and effective range.
For this we have taken advantage of a previous theoretical work that obtains a fair reproduction of the data using input of the chiral unitary approach.
We have relied on that approach, however allowing some flexibility on the parameters of the theory to obtain a perfect fit.
Since an important issue is the determination of the uncertainties, we have used the resampling method to do this job eliminating the problem on having to deal with the correlations between the parameters of the theory when such flexibility is allowed.
We obtain the values for the scattering length with uncertainties smaller that 20\% and the effective range is determined for the first time with a similar precision.
This result is most welcome, given the large dispersion of values provided in the literature for the scattering length so far.
Concerning the $a_0(980)$, we confirm what is becoming obvious, that that state does not have an ordinary pole in the second Riemann sheet and corresponds to a virtual state that shows up as a neat cusp in the $\eta\pi$ mass distribution.

\section*{Acknowledgments}
We thank Prof. Feng-Kun Guo for useful comment.
This work is partly supported by the National Natural Science Foundation of China (NSFC) under Grants No. 12365019 and No. 11975083, 
and by the Central Government Guidance Funds for Local Scientific and Technological Development, China (No. Guike ZY22096024), 
and by the Natural Science Foundation of Guangxi province under Grant No. 2023JJA110076.
This work is supported partly by the Spanish Ministerio de Ciencia e Innovaci\'on (MICINN) under contracts PID2020-112777GB-I00, PID2023-147458NB-C21 and CEX2023-001292-S; by Generalitat Valenciana under contracts PROMETEO/2020/023 and CIPROM/2023/59.

\bibliographystyle{a}
\bibliography{ref}
\end{document}